\documentclass[prl,preprint, superscriptaddress, showpacs]{revtex4}
\usepackage{graphicx} \usepackage{amssymb} \usepackage[usenames]{color}
\usepackage{amsmath}
\usepackage{xspace}

\newcommand{\ltsim}{\mbox{{\raisebox{-0.4ex}{$\stackrel{<}{{\scriptstyle\sim
}}
$}}}}

\begin{document}

\title{Dirty Peierls transitions in $\alpha$-Uranium}
\author{S. Cox}
\affiliation{National High Magnetic Field Laboratory, Ms-E536, Los Alamos, New Mexico, 87545, USA}
\author{E. Rosten}
\affiliation{Los Alamos National Laboratory, Los Alamos, New Mexico, 87545, USA}
\author{R.D. McDonald}
\affiliation{National High Magnetic Field Laboratory, Ms-E536, Los Alamos, New Mexico, 87545, USA}
\author{J. Singleton}
\affiliation{National High Magnetic Field Laboratory, Ms-E536, Los Alamos, New Mexico, 87545, USA}
\begin{abstract}
We point out that a recent model for the heat capacity of $\alpha$-U that invokes CDW collective
modes is unphysical.
We show instead that the features in the heat capacity of both single-crystal and polycrystalline
$\alpha$-U can be accounted for by a number of Peierls transitions that are subject to increased
disorder in the polycrystalline sample.
\end{abstract}


\maketitle
The interpretation of Uranium heat capacity reported in~\cite{alphaU}
requires the energy/temperature scales of the charge-density-wave (CDW) collective mode
(to which phenomena in the heat capacity are attributed)
and the CDW transitions themselves to be similar. However, for the collective mode to be a valid description
of the lowest-energy excited state of the CDW, the temperature should be such that
pair breaking is completely negligible; e.g. in the case of the CDW material 
(TaSe$_4$)$_2$I, the Peierls transition occurs at 263~K, whereas collective mode
contributions to the heat capacity are observed for $T~\ltsim ~1.7$~K~[2]. By contrast,
the authors of Ref.~[1] claim that collective mode contributions are responsible
for phenomena at and around the CDW ordering transitions. At such temperatures,
the single-particle gap is closing, and sub-gap excitations will be smeared out, 
invalidating the interpretation in Ref.~\cite{alphaU}.

\begin{figure}
\begin{centering}
\includegraphics[width=1.0\columnwidth]{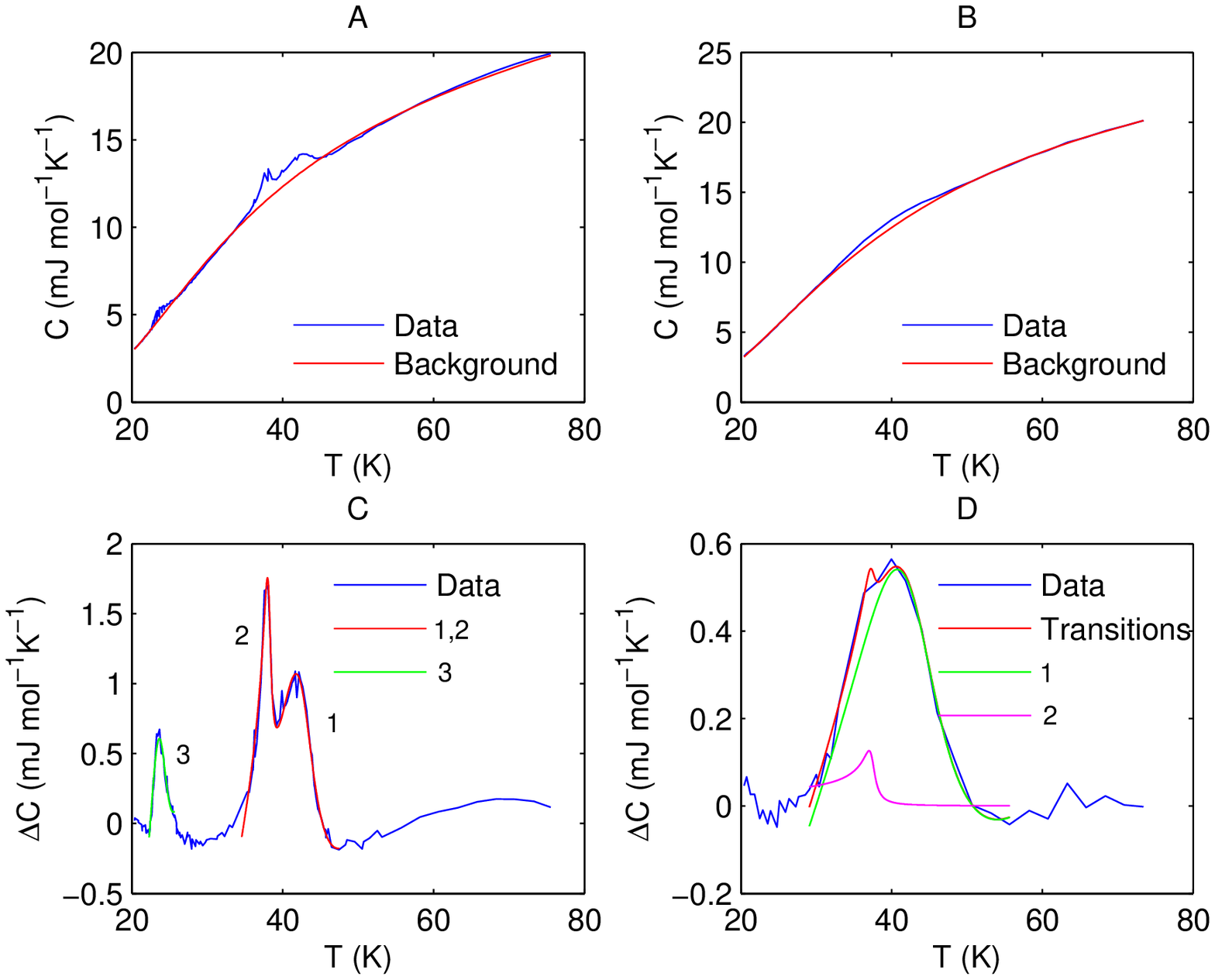}
\vspace{-5mm}
\caption{(a) Heat capacity of $\alpha$-U with modelled background and (c) heat capacity above background and dirty Peierls models 
for a single crystal. (b,d) Same quantities for a polycrystalline sample. 
\label{modelling}}
\vspace{-5mm}
\end{centering}
\end{figure}

A more compelling explanation of the specific heat anomalies  
reported in \cite{alphaU} is the model of Peierls transitions in a system with disorder or impurities 
(`dirty' Peierls transitions)~\cite{Chandra, dirty_me}.  In this model the transitions in the polycrystalline sample are broadened (or destroyed altogether)
by the increased level of disorder relative to the single crystal.  
In order to model the transitions the smooth background of the heat capacity data was removed~\cite{dirty_me}.  
Three transitions (numbered 1,2,3) were found in the single crystal as expected (Fig.~\ref{modelling}a and c).  In the 
polycrystalline sample an excess heat capacity was evident in 
the temperature range of transitions 1 and $2$ in the single crystal, with transition 3 absent 
(Fig.~\ref{modelling}b and d).
The model of a disordered Peierls transition fits the data well in both materials (Fig.~\ref{modelling}c and d), with the data
in the polycrystalline material modelled as two transitions.  
The fits to the transitions give a disorder lengthscale for each transition~\cite{Chandra, dirty_me}. 
The disorder lengthscales for the single crystal were 
$L_{1s}$=11.6 \AA, $L_{2s}$=25.8 \AA  ~and $L_{3s}$=12.3 \AA,  ~and for the polycrystalline sample were $L_{1p}$=8.3 \AA~
and $L_{2p}$=21.3 \AA.

Transition 1 is the onset of the three CDWs, all incommensurate.  At transition 2,
$q_x$ locks into $0.5a^*$, and at transition 3, $q_y$ locks into $\frac{1}{6}b^*$
and $q_z$ locks into $\frac{2}{11}c^*$~\cite{Grubel}.
Since the periodicity of the CDWs which are locking into the lattice at transition 3
are so large (6$b$ and 5.5$c$), it should be the 
transition most affected by disorder, explaining its disappearance in the polycrystalline sample.
This is consistent with observations in materials with similar ordering
in which the strain in a polycrystalline sample can prevent a lockin in some grains~\cite{philmag}.

Therefore the CDW transitions 
in both single- and polycrystalline $\alpha$-U can be modelled as Peierls transitions in a system containing different 
levels of disorder, without reference to pinning modes of the CDWs.

This research is supported by DoE grand LDRD-DR 20070013.  NHMFL is 
funded by NSF, DoE and the State of Florida. 
S.~Cox acknowledges support from the Seaborg Institute.
We thank P.B. Littlewood and N. Harrison for helpful comments.
\vspace{-3mm}

\bibliographystyle{apsrev}

\newpage

\end{document}